\documentclass[conference,10pt]{IEEEtran}
\usepackage{algorithm,multirow}
\usepackage[noend]{algorithmic}
\usepackage{graphicx}
\usepackage{comment}
\usepackage{caption}
\usepackage{multirow}
\usepackage{amssymb,amsthm,mathrsfs}
\usepackage{amsfonts,epsf,graphicx,times,amsmath,multirow,url,cite}
\usepackage{float}
\usepackage[capitalise]{cleveref}
\usepackage{mathtools}
\usepackage{amsmath}
\usepackage{array}
\usepackage{wrapfig}
\usepackage{tabularx}

\graphicspath{{figs/}}
	\DeclareGraphicsExtensions{.eps,.pdf,.jpeg,.png}

\makeatletter
\def\ps@headings{%
\def\@oddhead{\mbox{}\scriptsize\rightmark \hfil \thepage}%
\def\@evenhead{\scriptsize\thepage \hfil \leftmark\mbox{}}%
\def\@oddfoot{}%
\def\@evenfoot{}}
\makeatother
\newcommand{\sol}{\texttt{DRESS}}

\begin{document}

\date{}


\title{\huge DRESS: Dynamic RESource-reservation Scheme for Congested Data-intensive Computing Platforms}

\author{
\IEEEauthorblockN{
Ying Mao\IEEEauthorrefmark{1},
Victoria Green\IEEEauthorrefmark{1},
Jiayin Wang\IEEEauthorrefmark{2}, 
Haoyi Xiong\IEEEauthorrefmark{3},
Zhishan Guo\IEEEauthorrefmark{3} 
}
\IEEEauthorblockA{\IEEEauthorrefmark{1}Department of Computer Science,
The College of New Jersey, Email: \{maoy, greenv2\}@tcnj.edu}
\IEEEauthorblockA{\IEEEauthorrefmark{2} Department of Computer Science,
Montclair State University, Email: jiayin.wang@montclair.edu}
\IEEEauthorblockA{\IEEEauthorrefmark{3} Department of Computer Science,
Missouri University of Science and Technology, Email: \{xiongha,guozh\}@mst.edu}
}

\maketitle
\begin{abstract}
In the past few years, we have envisioned an increasing number of businesses start driving by big data analytics, 
such as Amazon recommendations and Google Advertisements.
At the back-end side, the businesses are powered by big data processing platforms to quickly extract information and make decisions.
Running on top of a computing cluster, those platforms utilize scheduling algorithms
to allocate resources.
An efficient scheduler is crucial to the system performance due to limited resources, e.g. CPU and Memory, and a
large number of user demands. 
However,
besides requests from clients and current status of the system, 
it has limited knowledge about execution length of the running jobs, and incoming jobs' resource demands, 
which make assigning resources a challenging task. 
If most of the resources are occupied by a long-running job, other jobs 
will have to keep waiting until it releases them.
This paper presents a new scheduling strategy, named \sol~that particularly aims to optimize the allocation
among jobs with various demands. Specifically, it classifies the jobs into two categories based on their requests,
reserves a portion of resources for each of category, and dynamically adjusts the reserved ratio by monitoring the pending requests and
estimating release patterns of running jobs. The results demonstrate \sol~significantly reduces the completion time
for one category, up to 76.1\% in our experiments, and in the meanwhile, maintains a stable overall system performance.

\end{abstract}
\section{Introduction}
Over the past few years, data-driven businesses have provided a promising experience for clients from various aspects.  
For example, Amazon provides personalized product recommendation using clients' past purchasing records, personal information (e.g., employment status and residence location) as well as the contextual information (e.g., weather).
%
%
On the other hand, 
Google's advertising system is optimized to retrieve the advertisements that the clients are potentially interested in using 
their browsing cookies, searching history, email contents, and even their friends' recent purchase records. 
Financial companies utilize machine learning ~\cite{yancy9175642, yancy2021form, zhu2021clustering, li2021frequentnet} and computational modeling~\cite{zhu2020high, jarrow2021low, zhu2021time, zhu2020adaptive, zhu2021news} to provide advanced services.
Given such tremendous data for processing, mining and analyzing, there needs a computing cluster that provides infrastructural supports for those data-intensive applications, at the back-end side of businesses.

To enable and optimize the big data analytics, the system usually first decompose the overall computational job into multiple small tasks, then itemizes the usage of systems into large set resources, such as CPU hour, or space of storage/memory, and further allocates computing resources to the tasks. 
%
%
%


To optimize the data processing systems, the researchers have put tremendous efforts on job scheduling, resource management, and program design to improve 
system performance.
In this field, there are two widely used schedulers, Fair~\cite{fair} and Capacity~\cite{capacity} schedulers, for managing the resources in the cluster.
Fair scheduler is a method of assigning resources to jobs such that all jobs get, on average, an equal share of resources over time. 
On the other hand, Capacity scheduler is designed to allow sharing a large cluster while giving each organization a minimum capacity guarantee.
Although these two schedulers have different strategies for resource management, 
both of them add jobs to the queues following a first-come-first-serve manner.

\begin{wrapfigure}{l}{0.23\textwidth}
\centering
\includegraphics[width=\linewidth]{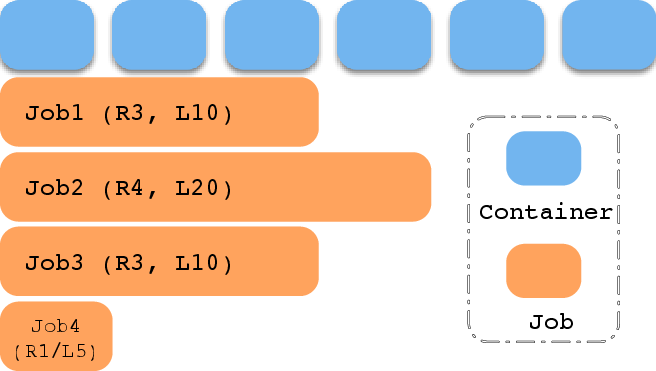}
\caption{4 incoming jobs for a cluster with 6 containers}
\label{fig:motivation-dress}  
\end{wrapfigure}
However, this manner does not take diverse demands from clients, which include various resource requests and occupancy length, into consideration.
Fig~\ref{fig:motivation-dress} illustrates a simplified  example of 4 incoming jobs in a cluster with 
6 containers. 
Suppose 4 jobs are submitted in order with a 1 second interval and each of them specifies its 
resource demand along with expected execution length (as shown on the figure). 
For example, Job 1 requests 3 (R3) containers and lasts for 10s (L10).
If the first-come-first-serve manner is in effect,
Job 1 executes first, followed by
Job 2, which waits until Job 1 finish  and finally, Job 3 \& 4 running in parallel. The makespan of four jobs would be 40s. 
The waiting time for Job 1 to Job 4 is 0s, 9s, 28s, 27s, respectively, and 16s on average.
In this schedule, only Job 3 and Job 4 are run in parallel in the system.
Towards a more efficient scheduler, it should consider the diverse demands from, not only running jobs, but also pending jobs.
In the above intuitive example, if the scheduler can delay the decision making and rearrange the execution order, where
Job 1 and 3 run concurrently, and then
Job 2 and 4 execute in parallel. 
Under the new schedule, the makespan reduces to 30s and average waiting time reduces to 5.75s.
Although Fig~\ref{fig:motivation-dress} is a simplified example, it shows the benefits that a system can obtain by considering 
the resource demands from clients.

Motivated by the fact that jobs with large demand will starve the system and delay other jobs, 
this paper presents \sol, a Dynamic RESource-reservation Scheme.
Compared to the prior work, \sol~collects the resource demands of jobs, distributes them into two categories with separate resource pools.
and, rearranges the execution order to increase the degree of parallelism. 
Specifically, \sol~utilizes a resource reservation ratio, which is calculated based on real-time demands, to allocate system resources
to each category. Additionally, \sol~estimates the resource release patterns. 
In summary, our main contributions are as follows:
 (1) We propose \sol~for congested data-intensive computing platforms that considers
 various demands from jobs and reserve resource for different categories.
 (2) We develop algorithms to estimate the resource release pattern for running jobs in the system. Based on the estimation along with 
 demands from waiting jobs, we dynamically adjust the resource reservation ratio.
 (3) We present a complete implementation on Hadoop YARN platforms. 
 The experiment-based evaluation 
 shows a significant 
 improvement on compilation time for small jobs, up to 76.1\%.

\vspace{-0.08in}
\section{Related Work}
Commercial companies and academic researchers utilize the computing systems to perform various jobs from different perspectives~\cite{chen2016overcoming, chen2017cyber,ding2016articulated, ding2015multilayer, chen2017secure, chen2017pinpointing, li2016deepcham, harvey2017edos, mao2017draps}.
There are many big data computing systems available in the market. Among various of them, the Hadoop YARN and its ecosystems have become major players.
Various applications from different users run on top the platforms and share resources in a cluster.
Traditionally, Fair~\cite{fair} and Capacity~\cite{capacity} are widely used to 
ensure each job to get a proper share of the available resources. 
To improve the performance of the computing systems, many research efforts have been spent on optimizing the system and job scheduling in different directions. Some major related works are introduced as follows. 

Resource-aware scheduling focuses on improving the resource utilization of the cluster. In this area, Haste~\cite{haste} is a fine-grained resource scheduling which leverages the information of requested resources and resource capacities to improve the resource utilization. 
In addition, FRESH~\cite{wang2014fresh} and OMO~\cite{wang2015omo} have developed dynamic resource management schemes according to the various workloads of different jobs. 
Another direction in system scheduling considers the heterogeneous environment. 
In this area, Teris~\cite{Tetris} packs tasks to machines based on their multiple resource requirements. LATE~\cite{LATE}, Hopper~\cite{Hopper}, and eSplash~\cite{esplash} aim to prevent unnecessary speculative executions in order to improve the performance in heterogeneous clusters. 

Data locality is also considered in the scheduling of big data computing systems. 
To improve the performance, authors in~\cite{suresh2014optimal} propose an optimal task selection algorithm
for better data locality and fairness. 
In addition, job characteristics are taken into consideration in the job-aware scheduling algorithms. In ARIA~\cite{aria}, a scheduler is proposed to allocate appropriate resources to jobs to meet the predefined deadline. 
Sparrow~\cite{Sparrow} targets on the scheduling problems with a huge amount of small jobs. 
Piranha~\cite{Piranha} creates an agent layer beyond Hadoop to schedule hybrid types of applications.  

Inspired by the preceding works, we develop a dynamic resource allocation scheme, {\sol}, to reserve a portion of resources for the applications with small resource requests. With a branch of hybrid jobs assigned in the cluster, based on the characteristics of each job and the estimating resources release of the cluster, {\sol} can significantly improve the performance of small jobs with limited impacts on large jobs. 

\section{\sol: Dynamic Resource Reservation Scheme}
\label{solution}
In this section, we present our solution \sol, which aims to reduce the waiting
time of small jobs in a congested cluster and at the same time, maintains a stable makespan among all jobs. The key idea of \sol~ is to redirect the 
jobs into two categories and reserve a certain amount of resources for each category.
As the system goes on, various jobs join and leave the categories when arriving and finishing.
The challenge lies in dynamically adjusting the reserved resources ratio to each category.
If a large portion of resources is reserved for one category, jobs in the other one would keep waiting.
Towards a better ratio adjustment, we not only need to know the total available resources, 
the number of pending jobs in each category,
but also the release patterns of running jobs to estimate the future availability of resources.
In the rest of this section, we first study the task execution in the parallel system,
then, describe our techniques to estimate the overall resource availability in the future.
Finally, based on the estimation, we propose an algorithm that dynamically adjusts the reserve ratio.
Table \ref{table:notations} lists the notations that are used in this paper. 
\vspace{-0.1in}
\begin{table}[ht]
\caption{Notation Table}
 \centering
 \small
 \begin{tabularx}{0.5\textwidth}{|c|X|}

  \hline 
  $J_i \in J$ & The $i^{th}$ job of all the jobs ($J$) in the cluster \\
  \hline
  $p_i \in J_x $ & The $i^{th}$ phase of a particular job ($J_x$)\\
  \hline
  $t_i \in p_x $ & The $i^{th}$ task of a particular phase ($p_x$)\\
  \hline
  $\alpha_i / \beta_i$ & The start and finish time of $J_i$ \\
  \hline
  $\gamma_{p_x}$ & The earliest finish time among all the tasks, $t_i \in p_x$ \\
  \hline
  $ps_{i_f} / ps_{i_l}$ & The starting time of the first / last task in $p_i$ \\  
  \hline
  $\Delta ps_i$ & The starting variation of $p_i$  \\
  \hline
  $f_i(t)/p_i(t)$ & Resource release function. Given $t$, it outputs an estimated number of containers that released by $J_i$/$p_i$ \\
  \hline
  $F(t)$ & Available resource function. Given $t$, it outputs an estimated number of available containers in system \\
  \hline
  $A_c/Tot_R$ & The number of available containers / total containers in the system  \\
  \hline
  $R_{J_i}/R_{p_x}$ & The number of containers occupied by $J_i$/$p_x$ \\
  \hline
  $RT(t)_{p_i} / CT(t)_{p_i}$ & The number of running/completed tasks in phase $p_i$ (represented by the states of containers) \\
  \hline
  $\delta$ / $\theta$ & Reserve ratio /  Job indicator \\
  \hline
\end{tabularx}
\label{table:notations}
\end{table}
\vspace{-0.2in}

\begin{figure*}[ht]
   \centering
      \begin{minipage}[t]{0.32\linewidth}
\centering
      \includegraphics[width=\linewidth]{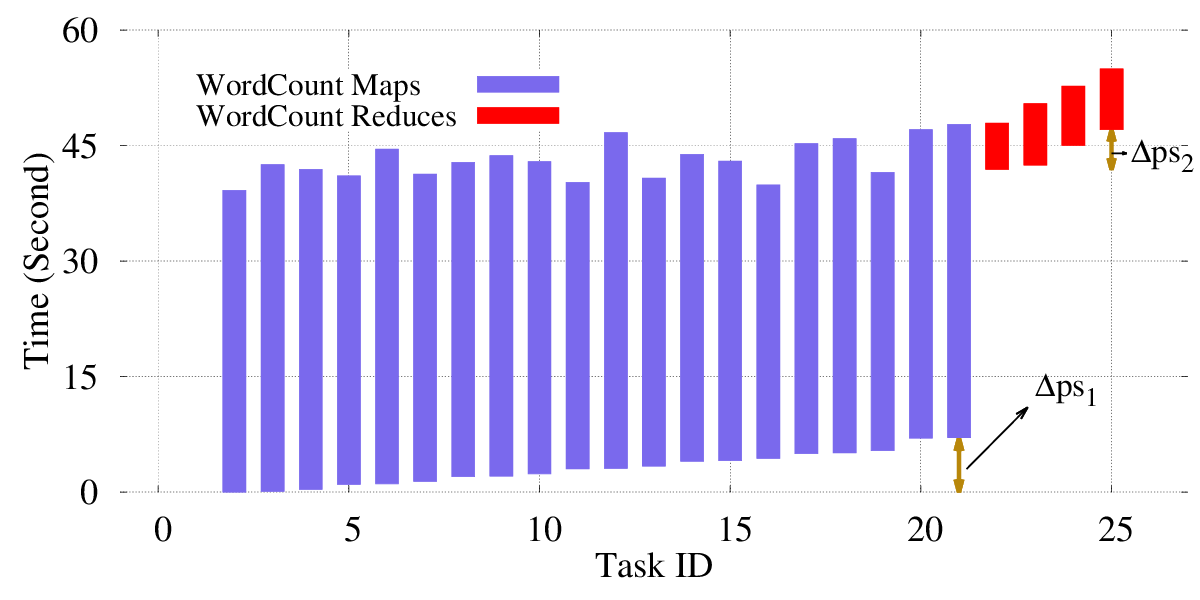}
\caption{A WordCount job on YARN with 20 Map Tasks and 4 Reduce Tasks (Starting Time Variation)}
      \label{fig:wc}
      \end{minipage} %
      ~
      \begin{minipage}[t]{0.32\linewidth}
\centering
      \includegraphics[width=\linewidth]{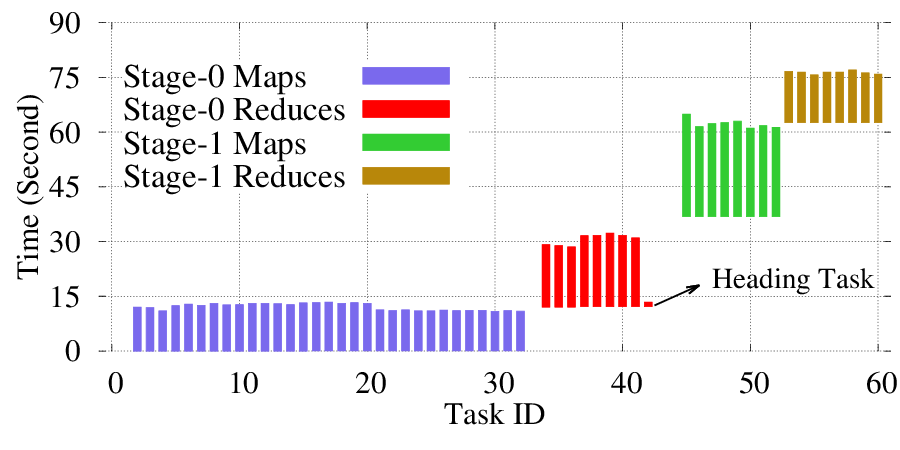}
\caption{A PageRank job with two MapReduce Stages on YARN (Heading Tasks)}
      \label{fig:pr}
      \end{minipage} %
      ~
      \begin{minipage}[t]{0.32\linewidth}
\centering
      \includegraphics[width=\linewidth]{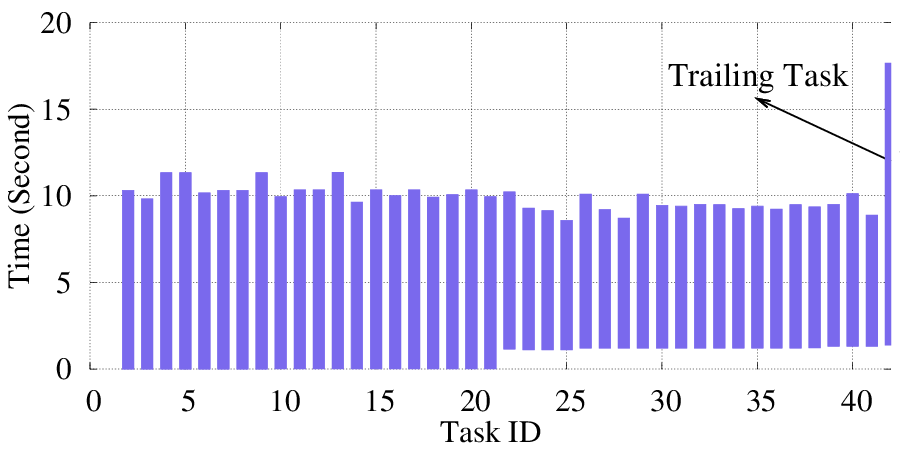}
\caption{A PageRank job with Spark-on-YARN (Trailing Tasks)}
      \label{fig:spark-pr}
      \end{minipage} %
\end{figure*}

\subsection{Characteristics of Task Execution}

Without prior knowledge about the features of data and algorithms, 
it's a challenge to estimate the job execution length. 
This is due to the fact that different jobs target on diverse data sets in terms of size, type, etc., 
and various algorithms will be applied to the data. For a better estimation, we 
experimentally study the task execution in three aspects, starting time variation, heading tasks, and trailing tasks.

\subsubsection{ Starting Time Variation} Focusing on a specific job, it consists of multiple tasks that can be grouped into multiple phases. 
Inside each phase, tasks perform the same operations with the same algorithms on similar data sets in order
to process it in parallel. 
Considering this characteristic, the task execution length in the
same phase should be similar with each other.  
Fig~\ref{fig:wc} plots a classic MapReduce WordCount job with 20 Map tasks and 4 Reduce tasks.
Clearly, the job contains two phases (Map and Reduce) and the tasks can be divided into two groups. 
As we can see the tasks in the same phase, Map and Reduce as on Fig~\ref{fig:wc}, have a similar execution length. 
The finishing time of Map tasks is varied due to the different starting time. 
There are mainly two reasons for the difference in starting time.
Firstly, in a congested cluster, 
the scheduler assigns the containers to jobs through multiple rounds of resource requests. Secondly, 
the transition delay varies from time to time when a container's state moves from {\bf New} to {\bf Running}, that passes by the other three states, {\bf Reserved}
{\bf Allocated}, and {\bf Acquired}. These reasons result in the starting time variances of jobs in each phase. 
As shown on Fig~\ref{fig:wc}, $\Delta_{ps_1}$ and $\Delta_{ps_2}$ for phase 1 and 2.

\subsubsection{ Heading Task} Fig~\ref{fig:pr} illustrates a PageRank example running on YARN with MapReduce. The PageRank job includes
two stages and each stage contains one Map and one Reduce phase. Therefore, tasks of a PageRank job can be naturally grouped into four phases.
It is clear to find the same trend of starting time variation on the tasks of the PageRank job. 
\begin{wrapfigure}{l}{0.30\textwidth}
\centering
\includegraphics[width=\linewidth]{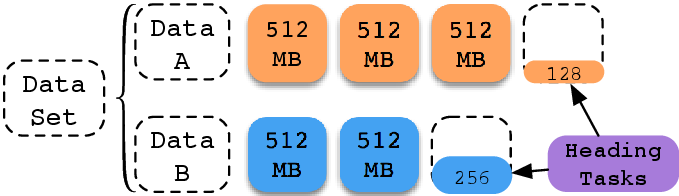}
\caption{Heading Tasks of a Job with two chunks in the dataset}
\label{fig:headingtasks} 
\end{wrapfigure}
However, there is an abnormal task in Reduce phase if the first stage that consists of 9 Reduce tasks.
While the average length of first 8 tasks is 18.25s with the variance of 1.45s, the last task (ID 42) only costs 1.26s that is less than 10\% of the others'.
This extreme case is caused by the fact that a large data set will be split into small data blocks and each task is responsible for
one or more blocks (controlled by the size of map split). 

Although the data blocks have the same limit on size, for tasks at the end, they may result in 
processing less data than previous tasks. Fig.~\ref{fig:headingtasks} shows an example of a job that targets on a data set of two chunks of 1,664MB (Data A) and 1,280MB (Data B).
The block size and map split are set to 512MB. Thus, Data A and B will be stored in four and three data blocks, respectively. The last blocks of Data A and B
are underloaded with only 128MB and 256MB, which lead to heading tasks.

\subsubsection{ Trailing Tasks} Fig~\ref{fig:spark-pr} plots a PageRank job running with Spark-on-YARN, which is a two-layer scheduling system, and we 
only collect data from Hadoop YARN. 
Unlike the previous heading task example, there is no distinct Map and Reduce phases on Spark.
In a Spark-on-YARN system, each task handles a partition of a large data set. However, due to the Data Skew Problem~\cite{kwon2013managing, cheng2014robust, cheng2014efficiently},
some partitions may much larger than others that lead to a longer execution time of those tasks, which we named trailing tasks.
We can easily locate one trailing task on Fig~\ref{fig:spark-pr} that costs 17.6s which consumes 38\% more than the second longest one.
Comparing with normal tasks, the trailing tasks occupy resources in a different pattern and significantly longer than others.


\subsection{Estimation Function}
The proposed solution, \sol, relies on an important parameter,
which is the estimated resource availability in the system. 
It is a critical factor that directly affects the estimated accuracy.
Utilizing characteristics in the previous subsection, we present our estimation function.
In our problem setting, we consider a system running multiple jobs simultaneously.
Each running job holds a certain amount of resources that are represented by containers.
A job is divided into multiple tasks and each of them runs in a container. 
Our objective is to estimate the overall resource availability of the system in the near future.

Suppose there are $n$ jobs, $J_1, J_2,...,J_n \in \{J\}$ in the system. For each one, $J_i \in \{J\}$, we define 
a function $f_i(t)$ to represent $J_i$'s estimated resource release frequency at time unit $t$. Let $F(t)$ denotes
the total number of available containers. Therefore, we have, 
\begin{multline}
 F(t) = F_1(t) + F_2(t) = A_c + f_1(t) + f_2(t) ... + f_n(t) 
 \label{totalC}
\end{multline}

Where $A_c$ is the number of currently available containers in the system that the scheduler can observe 
by monitoring the available resources on slave nodes, $F_1(t)$ and $F_2(t)$ are the estimated releasing resources for two categories.

For a specific $J_i$, before running, it does not occupy any containers. Thus, $f_i(t) = 0$, if $J_i$ is not started yet.
When $J_i$ finished all tasks, $f_i(t) = 0$ since all occupied resources have been released. Let $J_i$ starts at time unit 
$\alpha_i$ and finishes at $\beta_i$, where $\alpha_i$ and $\beta_i$ can be easily measured through heartbeat messages from slave nodes. 
The main activities, which including starting / executing tasks, occupying / releasing containers, and etc, 
happen inside the interval $[\alpha_i,~\beta_i]$. As described before, throughout the job execution, the tasks can be grouped into multiple phases that
tasks in each particular phase have the same operations, and similar input/output data for parallel processing. 
In an ideal setting, tasks in each phase will start and finish 
at the same time. However, in a real system, due to the limited resources and characteristics of task execution, the resource occupation and task completion
time varied in phases. Assume there are $m$ phases in the task execution of $J_i$ and let $p_j(t)$ 
be a function of resource release pattern for the $j^{th}$ phase in $J_i$,
then we have,

\begin{align}
f_i(t) = \left\{ \begin{array}{cl} 
                0 & \hspace{10mm} t < \alpha_i \\
                \sum_{0}^{m} p_j(t) & \hspace{10mm} t \in [\alpha_i,~\beta_i] \\
                0 & \hspace{10mm} t > \beta_i  \\
                \end{array} \right.
\label{equ:f}
\end{align}

Specifically, phase $j$, where $p_j \in J_i$, will not release any container until one of its task finishes.
When the first task finishes its execution, containers that occupied by this task will be returned to the system. 
Other tasks in phase $j$, which has the same operations and similar data sets, should about to complete. 
Depending on the starting time, the completion time of tasks in phase $j$
varies in a short period. We assume that the task completion time is equally distributed in the short period of $\Delta_t$,
where $\Delta_t$ can be measured from starting time variation. As a result, we have, 

\begin{align}
p_j(t) = \left\{ \begin{array}{cl} 
                0 & \hspace{10mm} t \in [\alpha_i,~\gamma_{p_j}] \\
                \frac{t - \gamma_{p_j}}{\Delta ps_j} \times c_{p_j} & \hspace{10mm} t \in [\gamma_i,~\gamma_{p_j}+\Delta ps_j] \\
                0 & \hspace{10mm} t \in [\gamma_{p_j} + \Delta ps_j,~\beta_i]  \\
                \end{array} \right.
\label{equ:p}
\end{align}
where $\alpha_i / \beta_i$ are the start and finish time of $J_i$ ($p_j \in J_i$);  
$c_{p_j}$ is the total number of containers that occupied by $p_j$; $\gamma_{p_j}$ is the earliest finishing time of the tasks in $p_j$;
$\frac{t - \gamma_{p_j}}{\Delta ps_j}$ is a percentage that represents the release progress in this phase.

\section{Parameter Analysis and Algorithms}
\label{algorithms}
From the analysis of the previous section, we can use Equation~\ref{totalC}, \ref{equ:f}, \ref{equ:p} to predict the resource release.
However, both $f_i(t)$ and $p_j(t)$ are based on several unknown parameters.
In this section, we analyze the parameters that are required by the equations and present the algorithms.

\subsection{Calculation of starting variation for each phase}
Calculating $f_i(t)$, we need to identify the phases for $J_i$ and determine the value of $\gamma_{p_j}$ and $\Delta ps_j$ for each $p_j \in J_i$.  
The key idea to estimate container release patterns of jobs is to group the task into different phases. 
Tasks of a phase run in parallel to achieve the same goal (e.g. producing intermediate output). 
Therefore, as the first step, we have to identify each phase of a job. As discussed in the previous section,
ideally, tasks in the same phase would start simultaneously. However, in reality, there is a starting time variation between them.
For each $p_j$ in $J_i$, we need to identify the start time of the first task in $p_j$ that denotes by $ps_{j_f}$ and start time of last task in $p_j$ that presents by
$ps_{j_l}$, which denoted by $\Delta ps_j = ps_{j_l} - ps_{j_f}$. 
We use a window-based algorithm to identify each phase.

\begin{algorithm}[ht]
\begin{algorithmic}[1]
\STATE $i=0$, $\alpha = 0$, $S_{p_j} =  false$, $c_{p_j}=0$
\STATE $ps_{j_f} = 0$, $ps_{j_l} = 0$, $R_{J_i} = \{R_{p_1}, R_{p_2}, ...\}$
\STATE $RT_{p_j}(t)$, $pw$: phase window 
\FOR {$t_i \in J_i$}
  \IF {$t_i.state = Running$}
 \STATE $J_i\rightarrow t_i$ and $R_{J_i} \leftarrow t_i$
  \STATE $RT_{p_j}(t) = |R_{J_i}|$ and $t_{i.start} = t$
  \STATE $c_{p_j} = c_{p_j} + 1$
      \IF {$i=0$}
      \STATE $\alpha_i = t$
      \ENDIF
    \IF {$RT_{p_j}(t) - RT_{p_j}(t-pw) > t_s$}
    \STATE $S_{p_j} = true$
    \STATE $ps_{j_f} = min \{t_{i.start}~|~R_{J_i}\}$
    \ELSIF {$ps_{j_f} \neq 0 $ and $RT_i(t) - RT_i(t-pw) = 0$ }
    \STATE $ps_{i_{l}} = max \{t_{i.start}~|~R_{J_i}\}$
    \STATE $\Delta ps_j = ps_{j_{l}} - ps_{j_{f}}$
    \STATE $i=i+1$
    \ENDIF
  \ENDIF  
\ENDFOR
\end{algorithmic}
\caption{Starting Variation of $j^{th}$ Phase for $J_i$}
\label{alg:start}
\end{algorithm}

Algorithm~\ref{alg:start} describes how we calculate $\Delta ps_j$ for $p_j \in J_i$. 
First of all, we initialize the parameters, where $j$ is a phase index, start time of $J_i$ that represents by $\alpha_i$, $S_{p_j}$ is a boolean indicator that is used
to determine whether $p_j$ has started, and  $ps_{j_f} ~/~ ps_{j_l}$ are initialized to 0.
$\{R_{J_i}\}$ is a set of running tasks for $J_i$ that grouped by each phases, and $RT_{p_j}(t)$ is a function that returns 
the number of running tasks in $p_j$ at system time$t$ (line 1-3).
The algorithm keeps tracking containers that have been assigned to $J_i$ but not in the ``Running'' state (may in Reserved, Allocated, and Acquired states).
If $t_i$ transits to ``Running'' state, which means the task is executing, it updates corresponding parameters (line 4-8).
In addition, we update the job starting time, $\alpha_i$, if $t_i$ is the first running task for $J_i$ (line 9-10).
Then, within a given window, the algorithm monitors the number of running tasks. If the difference is larger than a threshold($t_s$), 
it decides that the first task in $p_j$ has started at earliest starting time of $t_i$ in $\{R_{J_i}\}$ (line 11-13). 
On the other hand, if $RT_i(t)$ stays the same in the window,
the algorithm determines that the last task in $p_j$ has started at the latest starting time of $t_i$ in $R_{J_i}$,
calculates $\Delta ps_i$, and update the total number of running tasks in $p_j$ as well as the phase index $i$ (line 14-17).

\subsection{Calculation of starting release time of each phase}
Besides the $\Delta ps_j$, another key parameter for the estimation is $\gamma_j$, which is represented by the earliest finish time among tasks in $p_j$. 
According to equation~\ref{equ:p}, the phase $p_j$ will start releasing container in a period of
$\Delta ps_i$ after $\gamma_i$.

Algorithm~\ref{alg:end} presents a procedure to calculate the start releasing time, $\gamma_i$, of $p_j$ in $J_i$.
As the first step, it initializes the parameters, which are $\beta_i$ for $J_i$, the starting release time $\gamma_j$ for
$p_j$, $p_j$ ending indicator($E_{p_j}$),
running task set $R_{J_i}$ for $J_i$, completed task of $J_i$ that is represented by $CT_{J_i}$, 
and completed task function $CT_{p_j}(t)$ that returns the number of completed tasks for $p_j$ at time $t$ (line 1-3). 
The algorithm keeps tracking the states of running tasks in $R_{j_i}$, which are grouped by each phase.
When a container transits to ``Completed'' state, it indicates that the corresponding task, $t_i$ is finishing. The algorithm adds it into complete task' set, $CT_{J_i}$,
and updates $R_{J_i}$ as well as $CT_{p_j}(t)$ (line 4-7). In a given period ($pw$), if there are a certain number ($t_e$) of tasks move to 
``Completed'' state, it decides that tasks in $p_i$ have started finishing and records $\gamma_i$ (line 8-10). 
The threshold, $t_e$, is designed to filter out heading tasks. 
If $p_j$ has started finishing ($\gamma_i \neq 0$), but $CT_{p_j}(t)$ remain the same for a period ($pw$), additionally, $R_{J_i}$ for $p_j$ 
is not empty, it indicates that there are trailing tasks in $p_j$. In this case, we count trailing tasks into next phase (line 11-12).
Finally, if $R_{J_i}$ for $J_i$ becomes 0, all the tasks have finished and containers have been releases (line 13-14).
\begin{algorithm}[ht]
\begin{algorithmic}[1]
\STATE $\beta_i = 0$, $\gamma_j = 0$, $E_{p_j} = false$;
\STATE $R_{J_i} = \{R_{p_1}, R_{p_2}, ...\}$, $CT_{J_i} = \{CT_{p_1}, CT_{p_2}, ...\}$ 
\STATE $CT_{p_j}(t) = 0$, $t$: system time, $pw$: phase window

\FOR {$t_i \in R_{p_j}$}
  \IF {$t_i.state = Completed$}
  \STATE $t_i.finish = t$ and $CT_{p_j}(t) = |CT_{J_i}|$
  \STATE $CT_{J_i} \leftarrow t_i$ and $R_{J_i} \rightarrow t_i$
  \ENDIF
  
  \IF {$CT_{p_j}(t) - CT_{p_j}(t-pw) > t_e$}
  \STATE $E_{p_j} = true$
  \STATE $\gamma_j = min \{t_{i.finish}~|~C_i\}$
  
  \ELSIF {$\gamma_j \neq 0$ and \\ $CT_{p_i}(t) - CT_{p_i}(t-pw) = 0$ and $|R_{J_i}| > 0$}
  \STATE $c_{p_{i+1}} = c_{p_{i+1}} + |R_{p_i}|$
  
  \ELSIF {$|R_{J_i}|=0$}
  \STATE $\beta_i = t$
  \ENDIF
\ENDFOR
  
\end{algorithmic}
\caption{Starting Release Time for the $j^{th}$ phase of $J_i$}
\label{alg:end}
\end{algorithm}

With the values of $\alpha_i$, $\beta_i$ for $J_i$ and $\Delta ps_j$, and $\gamma_i$ for each phase $p_j \in J_i$, we could use the 
Equation~\ref{equ:f} and \ref{equ:p} to estimate the available containers in the system. 

\subsection{Dynamic configuration for reserved resource ratio}
In \sol, we estimate the available resources that guide the scheduler to dynamically adjust
the reserved resource ratio for each category. Besides available resources which can be predicted through Equation~\ref{totalC}, two more factors should
be taken into consideration, (1) the number of pending jobs in each category, (2) and the resource 
demands from them. Our objective of the dynamic configuration is
to reduce the average waiting and completion length, at the same time, maintain a stable overall system performance. 

While there are many approaches to split the jobs into different categories, such as job execution lengths, 
Map or Reduce intensive, and size of data sets, most of them require additional user-specified information. 
Requesting clients to input jobs' features need them fully understand both their jobs and targeted platforms, which
is not practical or feasible. In \sol, we use the resource demands of jobs as the indicator, which can be directly obtained
from the resource requests. 
We 
denote $\theta \in (0,1)$ as a preset indicator factor such that if the resource
request is larger than $A_c \times \theta$, the job will be classified to ``large demand''($LD$), otherwise, it will join ``small demand''($SD$).
In our problem setting, there are two categories in the cluster and we set $\theta = 10\%$ as the indicator. 
It's easy to classify incoming jobs into more categories by applying a similar strategy.

Each category maintains its own pool of jobs that consists of pending and running jobs. For pending jobs, 
the scheduler records the total demands of resources, in terms of containers.
For running jobs, the scheduler records the total occupied containers that will be returned to each category when the task completed.
The number of currently available containers, $A_c$, which can be observed from system heartbeats and 
can be further divided for each of the category, $A_{c_1}$ and $A_{c_2}$
, and $A_c = A_{c_1} + A_{c_2}$.  
Depending on jobs who release the containers, we can estimate available resource for each category with Equation~\ref{totalC}, where $F_1(t)$ and 
$F_2(t)$ are values for category 1 and 2.

With the parameters, \sol~ use Algorithm~\ref{alg:ratio} to dynamically adjust the reserved resource ratio, $\delta \in (0,1)$, 
which means $Tot_R \times \delta$ containers are assigned to ``small demand'' jobs,  $SD$, and $Tot_R \times (1-\delta)$ are for ``large demand'' jobs, $LD$,
where $Tot_R$ is the total number of containers in the system.
Firstly, we initialize the parameters (line 1-2). Then, the algorithm calculates total resource demands, $P_1$ and $P_2$ from all pending jobs in each
category(line 3-6). If, for $SD$, the estimated available resources are more than $P_1$ at time $t+1$, it assigns redundant resources
to $LD$ by reducing $\delta$ (line 7-8). On the other hand, if $P_1$ cannot be satisfied at time $t+1$ in $SD$, and, $LD$ has redundant resources,
it enlarges $\delta$ (line 9-11). If both $P_1$ and $P_2$ cannot be met by estimated resources, we sort the jobs in each category by their 
resource demands $r_i$, and start from the job with smallest resource demand, try to assign as many jobs as possible to utilize resources (line 12-20).
After the assignments, each of the categories may have some leftover ($A_{c_1}$ and $A_{c_2}$ may larger than 0). This is caused by the diversity in demands.
For example, if the smallest demand is 5 containers but the available resources are 4, these 4 containers are leftover. 
In this scenario, the algorithm tries to move the leftover from $LD$ to $SD$ since the jobs in $SD$ require fewer resources. 
Starting from the request of $J_{i+1}$, we check whether $r_{i+1}$ is less than the combined leftovers from $A_{c_1}$ and $A_{c_2}$.
It makes the maximum usage by checking $r_{i+1}$ iteratively until the next job request is larger than $A_{c_1} + A_{c_2}$.
The $\delta$ will be enlarged accordingly (line 21-24). Finally, the system will return the value of $\delta$ (line 25).

\begin{algorithm}[ht]
\begin{algorithmic}[1]
\STATE $SD$: Small demand jobs (category 1); $LD$: large demand jobs (category 2); $Tot_R$: total resources in the system;
\STATE $F_1(t+1)$/$F_2(t+1)$; $A_{c_1}$/$A_{c_2}$; $P_1/P_2$: resource demands from pending jobs in category 1 and 2;
\FOR {$J_i \in SD$}
\STATE $P_1 = P_1 + r_i$
\ENDFOR

\FOR {$J_i \in LD$}
\STATE $P_2 = P_2 + r_i$
\ENDFOR

\IF {$A_{c_1} + F_1(t+1) \geq P_1$}
\STATE $\delta  =  \delta - (A_{c_1}+F_1(t+1) - P_1) \div {Tot_R}$
\ELSE 
\IF {$A_{c_2} + F_2(t+1) \geq P_2$}
\STATE $\delta = \delta + (A_{c_2}+F_2(t+1) - P_2) \div {Tot_R} $
\ELSE
\STATE Sort in ascending order for $J_i \in SD$ and $J_j \in LD$ \\ based on $r_i$
\STATE $A_{c_1} = A_{c_1} + F_1(t+1)$ and $A_{c_2} = A_{c_2} + F_1(t+1)$
\FOR {$i=1;i<|SD|;i++$}
\IF {$A_{c_1} - r_i > 0$}
\STATE $A_{c_1} = A_{c_1} - r_i$
\ENDIF
\ENDFOR

\FOR {$j=1;j<|LD|;j++$}
\IF {$A_{c_2} - r_j > 0$}
\STATE $A_{c_2} = A_{c_2} - r_2$
\ENDIF
\ENDFOR

\FOR {$i=i+1;j<|SD|;i++$}
\IF {$r_i < A_{c_1} + A_{c_2}$}
\STATE $A_{c_2} = A_{c_2} - r_i$
\STATE $\delta = \delta + r_i \div Tot_R$
\ENDIF
\ENDFOR
\ENDIF
\ENDIF
\STATE Return $\delta$

\end{algorithmic}
\caption{Adjusting Reserve Resource Ratio}
\label{alg:ratio}
\end{algorithm}

\section{Evaluation}

\subsection{Implementation, Testbed and Workloads}

\subsubsection{Implementation, Testbed and Parameters}
We implement our solution \sol~on Hadoop YARN 2.7.4.
An enriched heartbeat message is used to transfer the required information, such as starting delays, between the master and slave nodes.
All the experiments are conducted on NSF Cloudlab~\cite{cloudlab} data center at University of Wisconsin.
We use the c220g2 server that has two Intel E5-2660 v3 10-core CPUs at 2.6 GHz (Haswell EP), 160 GB ECC Memory
and three disks (1 $\times$480 GB SATA SSD and 2$\times$ 1.2 TB HDDs ).

We launch a cluster with 5 nodes to evaluate \sol. As used by estimation functions, we set $t_s, t_e$ to 5s, phase window ($pw$)
to 10s, initial $\delta$ to 10\% and the job indicator $\theta=10\%$, such that large jobs request more than 10\%$\times A_c$.
Particularly, we choose a 5-node cluster, instead of a very large cluster, to simulate a congested
working environment for \sol. Moreover, due to the page limit, we omit the analysis of thresholds and phase window.

\subsubsection{Workloads}
To evaluate our system, we utilize a widely accepted benchmark suite named HiBench~\cite{hibench}.
In our settings, \sol~ can serve various types of jobs. There are 10 different benchmarks of 5
types, including micro benchmarks, machine learning, database, websearch benchmarks, and graph benchmarks.
Specifically, we have tested the following benchmarks:
(1) {\bf WordCount}: count the occurrence of each word in the input data, which are generated using RandomTextWriter~\cite{RandomTextWriter}.
(2) {\bf Sort}: sort its text input data, which is generated using RandomTextWriter.
(3) {\bf TeraSort}: sort (key,value) tuples on the key with the synthetic data as input.
(4) {\bf K-means clustering}: a well-known clustering algorithm for knowledge discovery and data mining and the input data set is 
generated by GenKMeansDataset~\cite{GenKMeansDataset}.
(5) {\bf Logistic Regression}: the Logistic Regression is implemented and the input data set is generated by LabeledPointDataGenerator~\cite{LabeledPointDataGenerator}.
(6) {\bf Bayesian Classification}: test the Naive Bayesian trainer with automatically generated documents whose words follow the zipfian distribution.
(7-8) {\bf Scan/Join}: SQL (Hive) queries.
(9) {\bf PageRank}: a search engine ranking benchmark.
(10) {\bf NWeight}: an iterative graph-parallel algorithm that computes associations between two vertices that are n-hop away.

Noted that streaming benchmarks are not included in the evaluation since they are long-running jobs and do not have resource
release patterns. 
In addition, we conduct the experiments on two type of platforms, Hadoop YARN (benchmarks 1-10) and 
Spark-on-YARN (benchmarks 4-6 and 9-10). 
When running the experiments, we consider 3 different combinations of jobs.
(1) {\bf MapReduce jobs}: we randomly pick up jobs for the Hadoop YARN platform and generate various sizes of datasets for each job.
(2) {\bf Spark jobs}: we randomly pick up the Spark jobs and execute them on Spark-on-YARN, 
which is a two-layer scheduling system (Spark has its own scheduler) and  \sol~only run on the YARN layer.
(3) {\bf Mixed job setting} : we randomly pick up jobs.
After selecting the jobs, they are submitted to the system one by one with a 5 seconds interval.

\begin{figure*}[ht]
   \centering
      \begin{minipage}[t]{0.4\linewidth}
\centering
      \includegraphics[width=\linewidth]{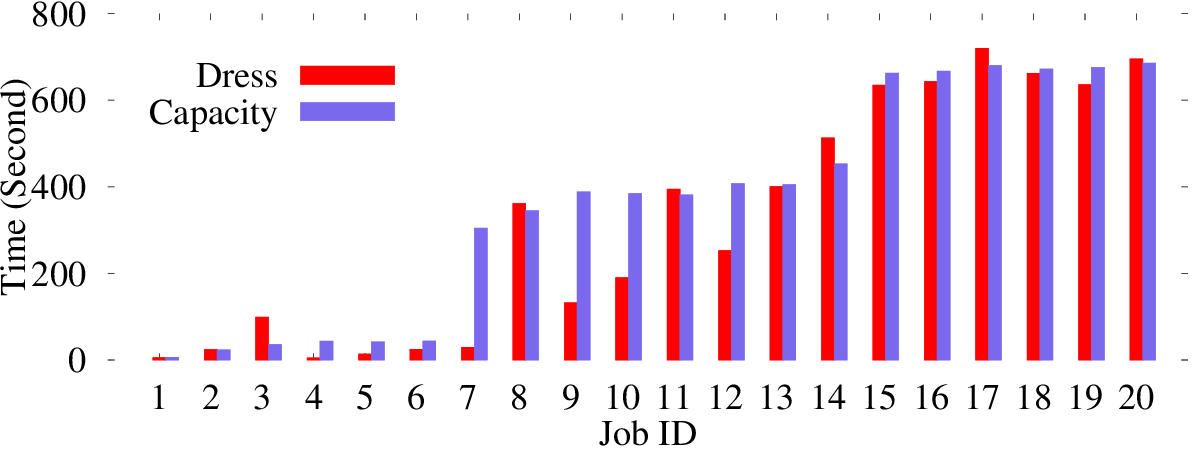}
\caption{Waiting Time of 20 Spark-on-Yarn Jobs}
      \label{fig:allspark:waiting}
      \end{minipage} 
      ~
            \begin{minipage}[t]{0.4\linewidth}
\centering
      \includegraphics[width=\linewidth]{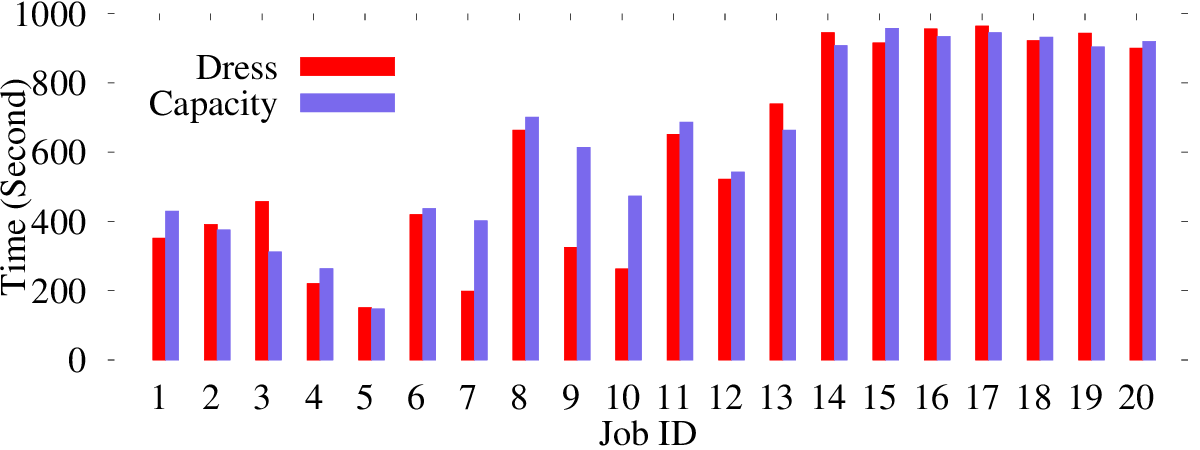}
\caption{Completion Time of 20 Spark-on-Yarn Jobs}
      \label{fig:allspark:length}
      \end{minipage} %
\end{figure*}

\begin{figure*}[ht]
   \centering
         \begin{minipage}[t]{0.4\linewidth}
\centering
         \includegraphics[width=\linewidth]{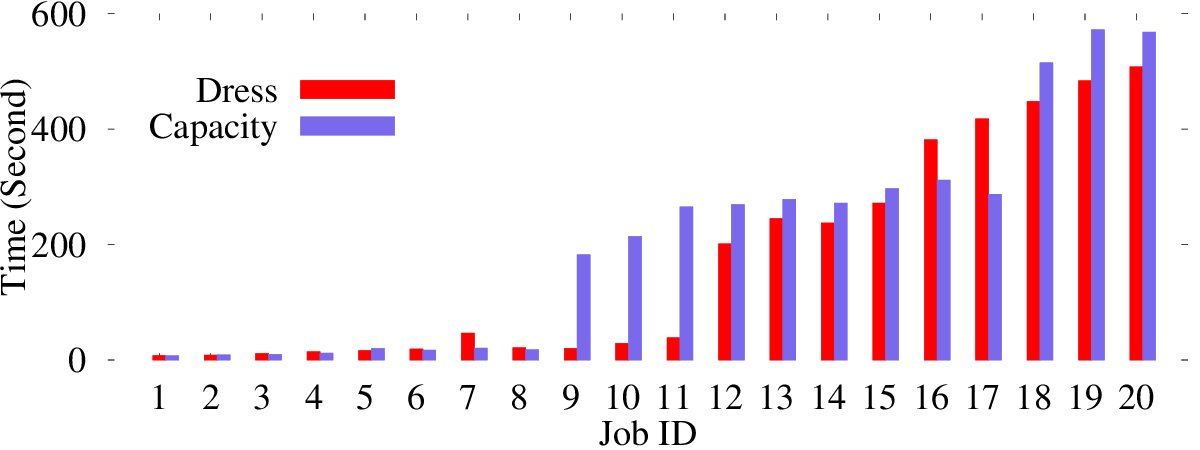}
\caption{Waiting Time of 20 MapReduce Jobs}
      \label{fig:allhadoop:waiting}
      \end{minipage} 
      ~
      \begin{minipage}[t]{0.4\linewidth}
\centering
      \includegraphics[width=\linewidth]{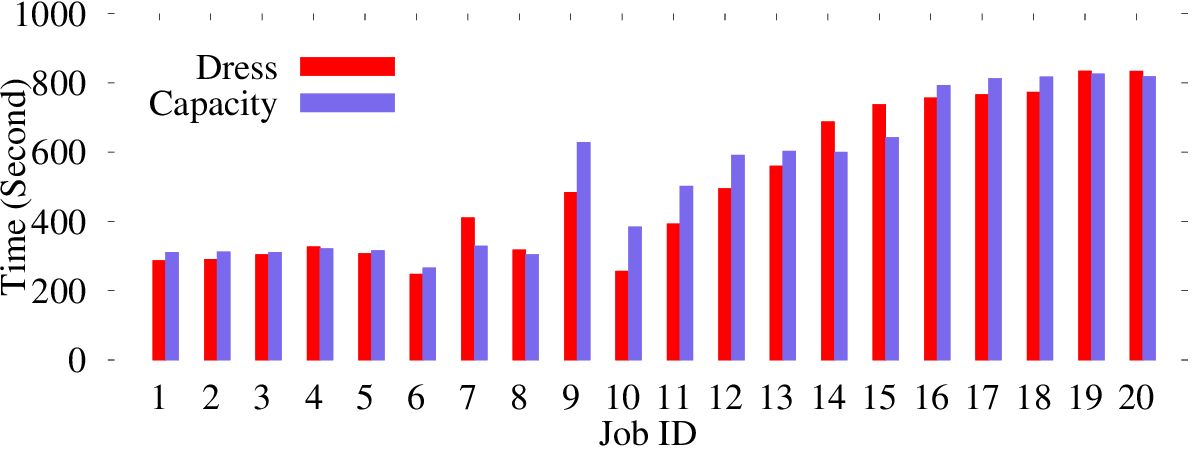}
\caption{Completion Time of 20 MapReduce Jobs}
      \label{fig:allhadoop:length}
      \end{minipage} %
\end{figure*}


\subsubsection{Evaluation Metrics} 
In the experiments, we mainly consider two performance metrics. 
From the system view, we compute the makespan that is the total execution time for all jobs. 
From the view of each individual job, we measure the waiting time and job completion time, 
e.g. from $J_i$, the waiting time
is the length from the submission of $J_i$ to the start of its first task, 
and the completion time is the length from the submission of $J_i$ to the completion of its
last task.
The makespan reflects an overall system performance across all the jobs, 
and the waiting time along with the completion time indicate how the system impact on each individual job.

\subsection{Experiment Results} 
\subsubsection{Spark-on-YARN}
Fig~\ref{fig:allspark:waiting} plots the waiting times of 20 jobs running on Spark-on-YARN, which contains a two-layer scheduling system, and
\sol~only runs on the YARN layer.
Overall, for the first 6 jobs running in the system, as the system is idle and the resources are enough to run the jobs in parallel, 
the waiting times of Jobs 1 to 6 are much shorter than others. 
Starting from Job 7 in Capacity scheduler, the waiting time for each job becomes higher than previous jobs.
This is because after running Job 1-6, the remaining resources in the system is not enough for Job 7 and it has to wait 
until one of them finish and release the resource.
Among tested jobs, ID 4, 5, 7, 9, 10, and 12 are jobs with small demands (less than 10 containers).
As illustrated in this figure, \sol can significantly reduce the waiting times of these small jobs compared to the Capacity scheduler. 
Especially, for \sol, the waiting time of Job 7 is more than 10x less than the one in Capacity scheduler (28.903s vs 304.705s). 
Since Job 7 is a small job, it can use the reserved resources and
after Job 4-5, the \sol~increased the reserved ratio to accept more small jobs. The same trend is found at Job 9, 10, and 12.
We also notice that the waiting time of Job 3 with \sol~is much longer than Capacity scheduler (98.863s vs 35.519s). After tracing back to the execution of jobs, 
we found that job 3 is delayed since a part of the resources are reserved for jobs with fewer demands. 
Fig~\ref{fig:allspark:length} illustrates the completion times of all 20 jobs. 
In \sol, for small jobs, the average reduction rate of the completion time is 27.6\%, with a maximum 51.2\% reduced completion time for Job 7.
We observe an increase on Job 3, 13, 14 of 32.0\%, 10.2\%, 6.1\%, and on average 16.1\%. 
In this case, the strategy of reserve resources does affect some jobs, however, it achieves a significant performance improvement on small jobs.
Table~\ref{table:spark} compares the makespan, average waiting time, average completion time as well as their median values of \sol and Capacity scheduler.
As illustrated in this table, the overall system performance, in terms of makespan, remains stable.

\vspace{-0.1in}
\begin{table}[ht]
	\centering
	\caption{Overall System Performance}
	\scalebox{0.95}{
		
		\begin{tabular}{ | c | c | c | c| c| |c | }				
			\hline
				     &   Makespan & Avg. W. & Median  & Avg. C. & Median \\ \hline	
			Capacity     &  1028.6	  &	310.1	 & 381.0 & 570.1 & 542.8 \\ \hline	
			\sol         &  1035.2    &     264.5    & 190.3 & 532.2 & 325.1 \\ \hline
		\end{tabular}	
	}
	
	\label{table:spark}
	\vspace{-5pt}
\end{table}


\subsubsection{Hadoop YARN}
Fig~\ref{fig:allhadoop:waiting} and Fig~\ref{fig:allhadoop:length} illustrate the results from experiments of 20 MapReduce jobs running on Hadoop YARN.
Comparing with the previous experiments, a similar trend can be discovered from Fig~\ref{fig:allhadoop:waiting}.
In 20 tested jobs, Job 4, 5, 6, 8, 10, 11 are jobs with small resource requests. 
As we can see from the figure, the waiting times for Jobs 1 to 9 are significantly shorter than others.
Unlike Job 3 in the Spark-on-YARN experiments, in Hadoop YARN tests, Job 7 has been delayed for the later small jobs (ID 8, 10,  and 11). 
In \sol, waiting time for Job 9 is much less than the same job running with Capacity (19.981s vs 189.246s). 
Although Job 9 is not a small job, it also gets benefit from the delayed running of Job 7 since unused resources 
not only distributes to the queue of small jobs, but also to the queue that targets on regular jobs. 

Fig~\ref{fig:allhadoop:length} presents the completion times for those 20 MapReduce jobs.
As the figure shows, \sol~reduces 25.7\%, on average, of completion times for small jobs.
In addition, it also benefits the large jobs of 9, 12, and 13. Their completion times decrease 23.2\%, 17.5\%, and 10.0\%, respectively.
\sol~sacrifices Job 7, whose completion time increased 29.3\%, and affects Job 14 and 15, which increased 12.2\% and 13.8\%.
Overall, in {\sol}, the completion times of 12 jobs are decreased by 18.5\% on average and the ones of other 8 jobs are increased by 8.2\% on average. 


\subsubsection{Mixed Job Setting}

Next, we present the results from a mixed job setting, where a cluster accepts both MapReduce and Spark jobs.
In addition, the number of jobs with small resource demands is another important for the system since it is directly
related to the dynamic configuration of reserved reservation ratio, which is controlled by Algorithm~\ref{alg:ratio}.

\begin{figure*}[ht]
   \centering
      \begin{minipage}[t]{0.40\linewidth}
\centering
      \includegraphics[width=\linewidth]{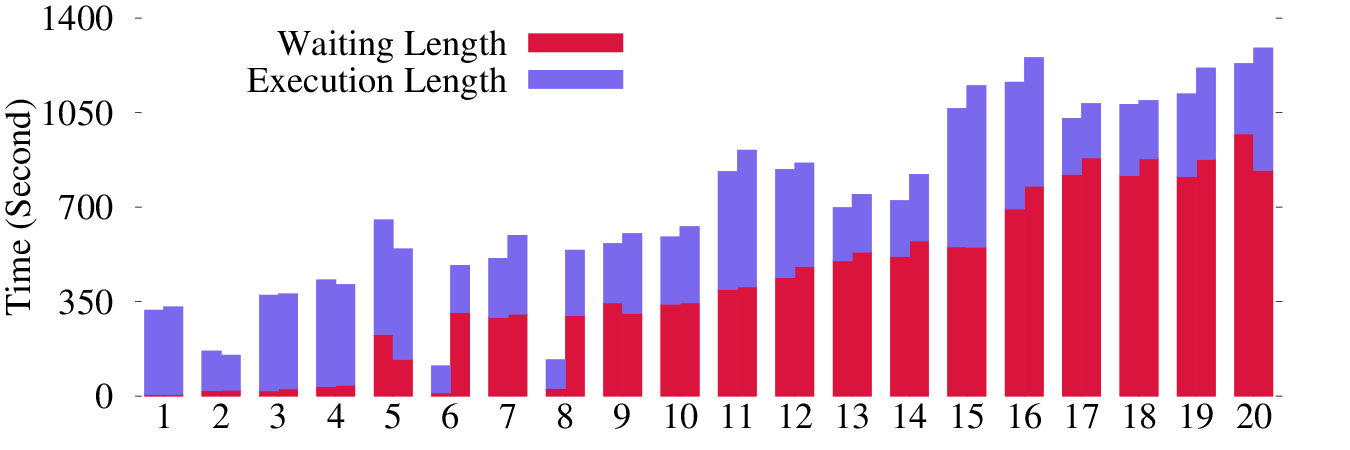}
\caption{Mixed Job Setting with 10\% Small Jobs}
      \label{fig:mix:1}
      \end{minipage} 
      ~
            \begin{minipage}[t]{0.40\linewidth}
\centering
      \includegraphics[width=\linewidth]{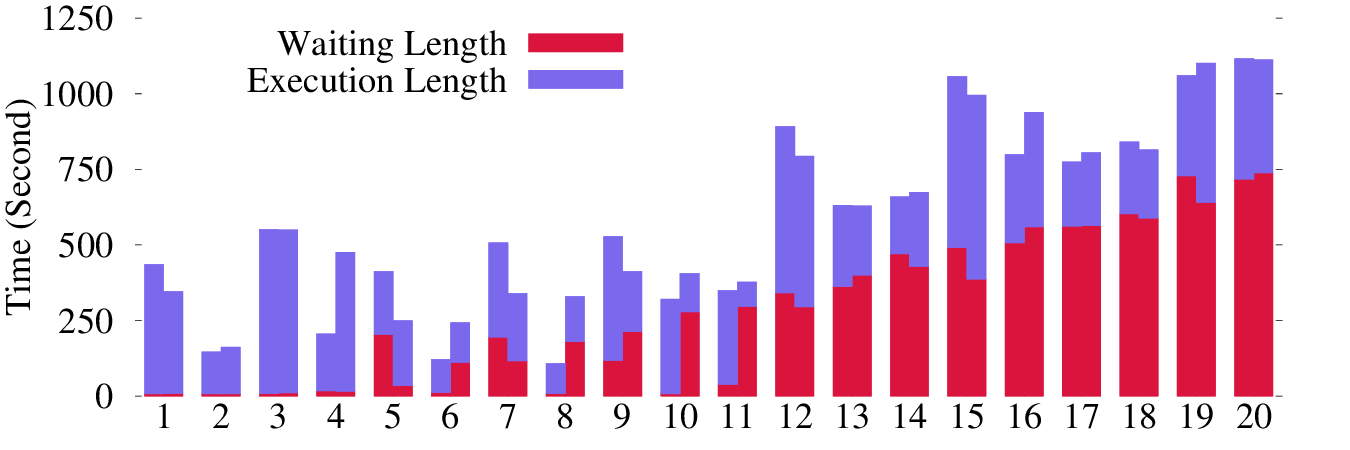}
\caption{Mixed Job Setting with 20\% Small Jobs}
      \label{fig:mix:2}
      \end{minipage} %
\end{figure*}
\begin{figure*}[ht]
\centering
      \begin{minipage}[t]{0.40\linewidth}
\centering
      \includegraphics[width=\linewidth]{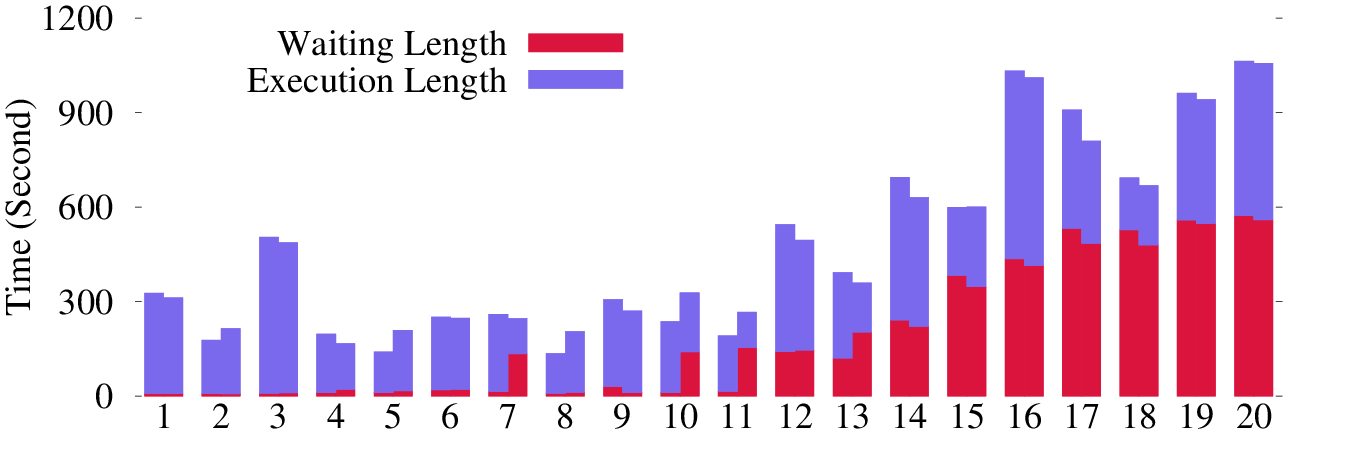}
\caption{Mixed Job Setting with 30\% Small Jobs}
      \label{fig:mix:3}
      \end{minipage} 
      ~
            \begin{minipage}[t]{0.40\linewidth}
\centering
      \includegraphics[width=\linewidth]{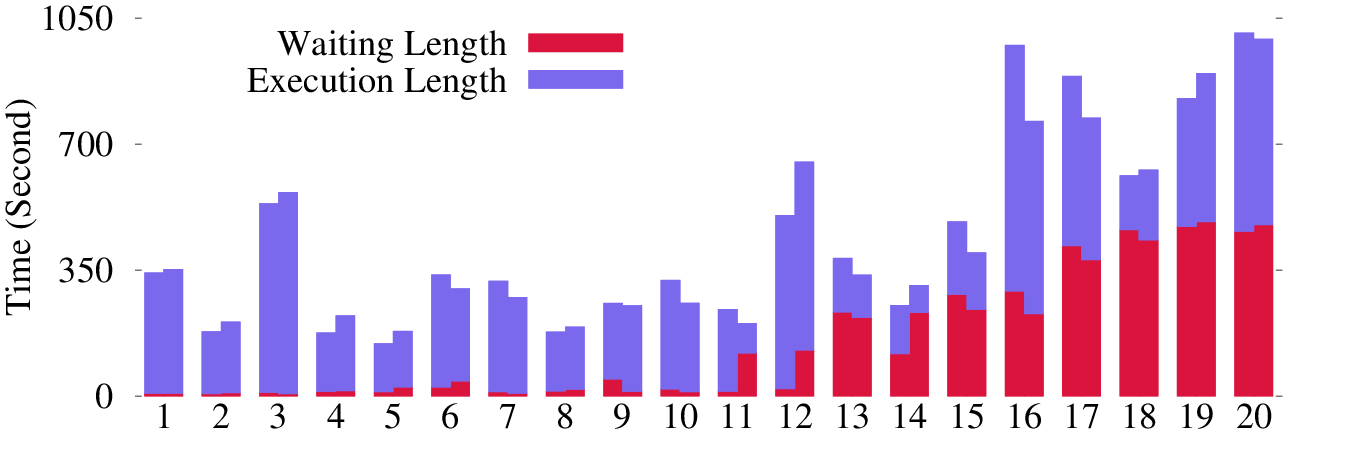}
\caption{Mixed Job Setting with 40\% Small Jobs}
      \label{fig:mix:4}
      \end{minipage} %
\end{figure*}

Fig~\ref{fig:mix:1},\ref{fig:mix:2},\ref{fig:mix:3},\ref{fig:mix:4} plot the experiments of mixed settings with 
10\%, 20\%, 30\%, and 40\% of small jobs. 
They plot the waiting time and the execution time for each job and the sum of them is the job completion time.
Two bars are illustrated for each job ID. The left bar shows the evaluation results of \sol~and the right bar shows the ones of Capacity scheduler. 
Through analyzing the data in the figures, we can derive a similar trend as we found from previous experiments.
Overall, the completion time of small jobs is significantly reduced in \sol~compared to Capacity scheduler. 
For instance, 
the completion times of Job 6 and 8, have been reduced 
from 484.5s, 540.3s to 111.3s, 134.4s, which is decreased by 76.1\% on average.
The reductions of the completion time for small jobs in the other three job settings are 36.2\%, 21.9\%, and 23.7\% on average.

\section{Conclusion}
This paper investigates the resource management in congested clusters.
Our goal is to reduce the waiting time and improve the completion time for jobs with fewer resource requests. 
To achieve our objectives, we present \sol, a dynamic resource reservation scheme.
Specifically, depending on incoming jobs' demands, 
it categories them into two categories.
\sol~reserves a portion of resources for jobs with small resource requests. The reserve ratio can be dynamically
adjusted according to the number of jobs in each queue.
We implement \sol~in the Hadoop YARN and evaluate it with both MapReduce and Spark jobs.
The experiment result shows a significant improvement in small jobs, up to 76.1\% reduction on the average completion time, 
in the meanwhile, achieves a stable overall system performance.
\bibliographystyle{unsrt}
\bibliography{routing}
\end{document}